\documentclass[10pt,letterpaper,compsoc,conference]{iiswc26}

\usepackage{cite}
\usepackage{amsmath,amssymb,amsfonts}
\usepackage{algorithmic}
\usepackage{graphicx}
\usepackage[dvipsnames]{xcolor}
\usepackage[final]{microtype}
\usepackage[italic]{mathastext}
\usepackage{libertine}
\usepackage[T1]{fontenc}
\usepackage{textcomp}
\usepackage[varqu,varl]{zi4}
\usepackage[all]{nowidow}
\usepackage[keeplastbox]{flushend}
\usepackage{fancyhdr}
\usepackage{listings}
\usepackage{booktabs}
\usepackage{xcolor}
\usepackage{colortbl}
\usepackage{multirow}
\usepackage{pifont}

\usepackage{xcolor}
\usepackage{colortbl}
\usepackage{multirow}
\usepackage{booktabs} 
\usepackage{tcolorbox}
\tcbuselibrary{skins}

\usepackage{subcaption}
\usepackage[dvipsnames,svgnames,x11names]{xcolor}
\newtcbox{\myyellow}[1][yellow!30]{ 
  on line, 
  colback=#1, 
  colframe=#1, 
  boxsep=0pt, 
  left=0pt, right=0pt, 
  top=0.2pt, bottom=0.2pt, 
  arc=2pt, 
  outer arc=2pt
}

\newtcbox{\mygreen}[1][YellowGreen!30]{ 
  on line, 
  colback=#1, 
  colframe=#1, 
  boxsep=0pt, 
  left=0pt, right=0pt, 
  top=0.2pt, bottom=0.2pt, 
  arc=2pt, 
  outer arc=2pt
}
\newtcbox{\myblue}[1][RoyalBlue!70]{ 
  on line, 
  colback=#1, 
  colframe=#1, 
  coltext=white,
  fontupper=\bfseries,
  boxsep=0pt, 
  left=0pt, right=0pt, 
  top=0.2pt, bottom=0.2pt, 
  arc=2pt, 
  outer arc=2pt
}

\newtcbox{\myorange}[1][orange!20]{ 
  on line, 
  colback=#1, 
  colframe=#1, 
  boxsep=0pt, 
  left=0pt, right=0pt, 
  top=0.2pt, bottom=0.2pt, 
  arc=2pt, 
  outer arc=2pt
}

\usepackage{tcolorbox}
\usepackage{enumitem}
\tcbuselibrary{skins} 
\definecolor{myBlue}{HTML}{8FABDB}
\usepackage{hyperref}
\newtcolorbox{observation}{
    enhanced,
    colback=myBlue!18,      
    colframe=myBlue!18,     
    arc=4pt,                
    boxrule=0pt,            
    left=6pt,               
    right=6pt,              
    top=4pt,                
    bottom=4pt,             
    before skip=4pt,        
    after skip=4pt,         
    fontupper=\linespread{1.1}\selectfont
}

\usepackage{tikz}
\newcommand{\circlednum}[1]{%
	\tikz[baseline=(char.base)]\node[shape=circle, draw=black, fill=black, text=white, inner sep=0.5pt, minimum size=2pt, font=\sffamily\bfseries\footnotesize] (char) {#1};%
}

\begin{document}

\IEEEoverridecommandlockouts


\title{Tools-CC-Bench: a Benchmark Suite for Collective Communication with Compression in HPC and AI Workloads}




\author{
\IEEEauthorblockN{Haozhe Fan\IEEEauthorrefmark{1}\IEEEauthorrefmark{3}\IEEEauthorrefmark{4}, Wei Wang\IEEEauthorrefmark{2}\IEEEauthorrefmark{4}, Xingchen Liu\IEEEauthorrefmark{1}, Man Liu\IEEEauthorrefmark{1}, Xingjian Tian\IEEEauthorrefmark{1},\\
Haoquan Long\IEEEauthorrefmark{1}, Zedong Liu\IEEEauthorrefmark{1}, Daran Sun\IEEEauthorrefmark{1}, Jinwu Yang\IEEEauthorrefmark{1}, Bo Yang\IEEEauthorrefmark{2},\\
Jie Liu\IEEEauthorrefmark{2}, Yonggang Che\IEEEauthorrefmark{2}, Hairui Zhao\IEEEauthorrefmark{1}, Guangming Tan\IEEEauthorrefmark{1}, Dingwen Tao\IEEEauthorrefmark{1}\IEEEauthorrefmark{5}}
\IEEEauthorblockA{
\IEEEauthorrefmark{1}Institute of Computing Technology, Chinese Academy of Sciences, Beijing, China\\
\{liuxingchen232, liuman24, tianxingjian25, longhaoquan25\}@mails.ucas.ac.cn\\
\{sundaran24s, yangjinwu24z, zhaohairui, tgm, taodingwen\}@ict.ac.cn, \{captain.liu77\}@gmail.com\\
\IEEEauthorrefmark{2}College of Computer Science and Technology, National University of Defense Technology, Changsha, China\\
\{weiwang\_jsjxy, yb, liujie, ygche\}@nudt.edu.cn\\
\IEEEauthorrefmark{3}School of Computer Science, Nanjing University, Nanjing, China\\
\{231180018\}@smail.nju.edu.cn
}
\thanks{\IEEEauthorrefmark{4}These authors contributed equally to this work.}
\thanks{\IEEEauthorrefmark{5}Corresponding author: Dingwen Tao, taodingwen@ict.ac.cn.}
}

\maketitle
\pagestyle{empty}


\begin{abstract}
Distributed HPC and LLM workloads increasingly require efficient communication for scalability, yet growing data movement has become a major performance bottleneck. Communication compression can reduce this overhead and complement execution-level optimizations, but its benefits remain difficult to assess because existing benchmarks lack support for diverse backends, realistic datasets, application-specific accuracy metrics, and overlap-induced resource contention. We present CC-Bench, a lightweight, extensible, and application-oriented benchmark suite for evaluating communication compression under realistic execution conditions. CC-Bench uses declarative application-environment modeling to decouple profiling logic from communication libraries, datasets, and fidelity metrics, enabling portable cross-library evaluation. It further combines function-level interception and hardware counter monitoring to characterize per-phase latency, hardware utilization, numerical fidelity, and computation interference. With representative datasets from HPC and LLM workloads, CC-Bench evaluates three compression-enabled communication libraries on CPU and GPU clusters, revealing accuracy-performance trade-offs and bottlenecks to guide practical deployment and optimization.

\end{abstract}

\section{Introduction}

Modern HPC and LLM workloads increasingly exceed the capacity of a single device. Applications ranging from molecular dynamics simulations with tens of billions of atoms to AI models with hundreds of billions of parameters rely on distributed systems to scale computation across many devices~\cite{pravsnikar2024machine,liang2024communication,Jiang2024MegaScaleSL,Cao2023,touvron2023llama,Brown2020gpt3, PaLM, Smith2022UsingDA}. Communication is fundamental to this scaling, enabling data exchange and synchronization. However, communication demand grows rapidly with workload scale, while network bandwidth remains comparatively constrained. Consequently, communication overhead has become a major bottleneck for workload scalability and end-to-end performance~\cite{narayanan2021efficient}.


Existing efforts to mitigate communication overhead mainly fall into two categories. One line of work improves communication execution efficiency through topology-aware optimization~\cite{kim2024tccl, wang2023topoopt, shah2023taccl
}, optimized communication algorithms~\cite{cai2021synthesizing, cowan2023mscclang, de2024swing}, and communication resource management~\cite{zhang2025comet, chen2024centauri, cheng2025concerto, jangda2022breaking, wang2022overlap}, facilitated by mature profiling and analysis tools~\cite{lee2024collective,Hanpeng2022,gu2026ccld}. Another line reduces the volume of transmitted data through compression. Communication compression is attractive because it is broadly applicable, offers substantial performance benefits, and complements execution-level optimizations. It has therefore become increasingly effective in real-world applications~\cite{jia2024sdp4bit,he2025tah,liu2026taco}, and is now supported by a growing number of communication libraries~\cite{liucoccl2026,lin2026zipccl,ma2026ucclzip,wang2026ncclz}. However, despite its promise and growing research interest, communication compression remains difficult to evaluate in realistic workloads due to the lack of efficient benchmarking tools. As a result, its practicality and performance characteristics remain poorly understood, limiting effective optimization and large-scale deployment for communication-intensive workloads.

\begin{figure}[t]
    \centering
    \includegraphics[width = 1.0\linewidth]{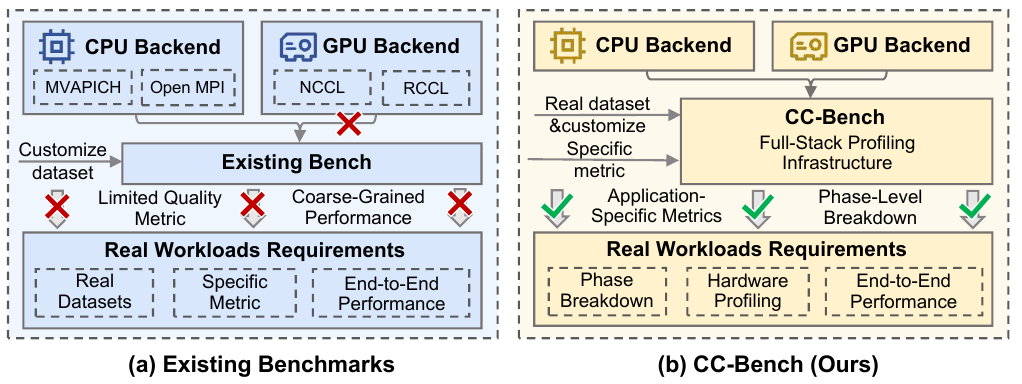}
    \caption{Benchmarking compression-supported communication libraries using our bench and previous methods.}
    \label{fig:exsiting vs ours}
\end{figure}

Traditional communication benchmarks, such as OSU Micro-Benchmarks and XCCL-Tests~\cite{panda2021mvapich,shroff2000collmark}, primarily measure communication efficiency but provide limited support for communication compression. Evaluating compression in real applications requires jointly characterizing compression ratio, throughput, application-specific accuracy, and interference with overlapped computation. Although recent efforts have added compression support, they remain limited in three aspects~\cite{ombcompr2025}, as shown in Fig.~\ref{fig:exsiting vs ours}(a). First, they are largely MPI-only and difficult to extend, failing to cover the diverse communication libraries used by modern applications. Second, communication compression is data-dependent: compression ratio and numerical fidelity vary across datasets, while accuracy requirements differ across applications. However, current benchmarks provide limited datasets and accuracy metrics, making it difficult to assess compression methods across workloads. Third, compression may contend with application computation during overlap, while existing benchmarks only report coarse-grained latency and throughput for isolated communication compression. This limits their ability to reveal end-to-end speedups and performance bottlenecks in real applications.

These mismatches motivate lightweight, extensible, and application-oriented benchmarks that evaluate communication compression under realistic execution conditions without requiring full application deployment. Developing such a benchmark suite, however, introduces several challenges: 1) Heterogeneous communication backends across applications hinder unified cross-library evaluation. 2) Opaque communication compression implementations complicate fine-grained performance attribution. 3) Diverse application data characteristics, accuracy requirements, and execution patterns render fixed benchmark configurations insufficient for realistic end-to-end workloads.

To address these challenges, we present \textbf{CC-Bench\footnote{https://anonymous.4open.science/status/CC-Bench}}, a lightweight, customizable, and application-oriented benchmark suite for evaluating communication compression across diverse communication backends. As shown in Fig.~\ref{fig:exsiting vs ours}(b), CC-Bench enables systematic communication compression analysis through application-oriented configuration and runtime performance profiling, enabling realistic workload emulation during benchmark execution. In addition, CC-Bench incorporates communication datasets from representative applications, enabling flash cross-domain evaluation of communication compression. To the best of our knowledge, CC-Bench is the first \textbf{open-source} benchmark suite for comprehensive evaluation of communication compression. Our contributions are summarized as follows:

\begin{itemize}[leftmargin=*, noitemsep, topsep=0pt]

    \item We introduce an \textbf{application-oriented configuration module} that decouples profiling logic from communication libraries, datasets, and error metrics through unified abstractions, supports user-defined integration of customized components, and enables declarative configuration to substantially simplify cross-application evaluation.


    \item We develop a \textbf{full-stack profiling module} that combines function-level interception with hardware counter monitoring for systematic characterization of communication compression. The framework captures per-phase latency, records function traces and monitors hardware utilization under varying resource interference, exposes performance breakdowns and overlap-induced contention, and quantifies numerical fidelity across datasets.


\item We incorporate \textbf{built-in representative datasets} spanning scientific computing and LLM training/inference, enabling convenient evaluation of communication compression methods across diverse workloads.
    
     \item We use CC-Bench to \textbf{evaluate three compression-supported communication libraries} on CPU and GPU clusters across HPC and AI workloads. The evaluation characterizes accuracy, performance trade-offs, and overhead breakdowns under diverse computational conditions, providing practical deployment and optimization insights. These results show that CC-Bench bridges basic benchmark evaluation and full-application profiling for communication compression.

\end{itemize}

The remainder of this paper is organized as follows. Sec.~\ref{sec:background} reviews communication workloads, libraries, compression techniques, and benchmark gaps. Sec.~\ref{sec:methodology}  presents the fundamental goals and design principles of our benchmark. Sec.~\ref{sec:design} illustrates the architecture, workflow and coverage of CC-Bench along with a usage demonstration.  Sec.~\ref{sec:evaluation} presents evaluations of compression-supported communication libraries on both CPU and GPU HPC clusters, and Sec.~\ref{sec:conclusion} discusses future directions.

\section{Background} \label{sec:background}

This section covers communication in HPC and AI workloads (Sec.~\ref{subsec:com_bg_ai}), collective communication libraries (Sec.~\ref{subsec:bg_comm_libs}), compression techniques (Sec.~\ref{subsec:comm_compr}), and the limitations of existing benchmarks (Sec.~\ref{subsec: limitations}).

\subsection{Communication for HPC and AI Workloads} \label{subsec:com_bg_ai}

As application scales continue to grow, a single device is increasingly insufficient to sustain modern workloads. Distributed execution therefore relies on inter-device communication to coordinate computation and synchronize intermediate states across multiple devices~\cite{ben_nun_2019}. Two mainstream classes of distributed workloads—scientific computing and AI—exhibit markedly different communication patterns. In distributed scientific computing, applications traditionally run on CPU clusters using process-based parallelism, where CPU-side collective primitives such as AllGather, AllReduce, and ReduceScatter are used to exchange intermediate results and global states~\cite{mpi_forum}. These communication operations are typically coarse-grained, with each transfer ranging from hundreds of megabytes to tens of gigabytes~\cite{hoefler_2011}. Recently, HPC workloads have increasingly offloaded computation to GPUs, leading to heterogeneous CPU–GPU execution that requires runtime switching between CPU- and GPU-based communication paths~\cite{wang_2011_mvapich2gpu}.

AI workloads, including AI-for-Science, large-scale model training, and inference services, are primarily executed on GPU clusters and employ multiple parallelization strategies, such as data parallelism (DP), tensor parallelism (TP), and pipeline parallelism (PP)~\cite{shoeybi2019megatronlm}. Different strategies induce distinct communication patterns. TP mainly relies on intra-node AllReduce to aggregate partial activations, while PP uses point-to-point communication to transfer activations across pipeline stages on different nodes. These communications are often only tens of megabytes per operation, but can occur thousands of times within a single iteration. In contrast, DP primarily uses ReduceScatter and AllGather to synchronize gradients and model parameters, where individual transfers can reach several to tens of gigabytes but occur only a few times per iteration.

\subsection{Collective Communication Libraries} \label{subsec:bg_comm_libs}

To support increasingly complex communication patterns, distributed applications rely on collective communication libraries that are tightly coupled with the underlying hardware platform. In high-performance scientific computing, where workloads primarily execute on CPU clusters, the Message Passing Interface (MPI) remains the de facto communication standard~\cite{mpi_forum}. As scientific computing workloads are increasingly offloaded to GPUs, CUDA-aware MPI has become widely adopted to enable direct GPU memory access and efficient communication across heterogeneous CPU–GPU systems~\cite{wang_2011_mvapich2gpu}. MPI provides a rich set of collective communication primitives, including AllReduce, ReduceScatter, AllGather, AlltoAll, and their vector variants, organized through the communicator abstraction. Modern MPI implementations, such as MVAPICH and Open MPI~\cite{panda2021mvapich, gabriel_2004,mvapich_project}, map these primitives onto high-performance interconnects (e.g., InfiniBand, OmniPath, and Ethernet) using topology- and message-aware communication algorithms, while also supporting CUDA-aware extensions.

Modern AI workloads, including AI for Science, LLM training, and inference, predominantly execute on GPU clusters and rely on GPU-specific collective communication libraries to exchange data efficiently across GPUs. Representative libraries include the NVIDIA Collective Communication Library (NCCL)~\cite{nccl} and the ROCm Communication Collectives Library (RCCL)~\cite{rccl}. NCCL provides optimized collective primitives for multi-GPU and multi-node communication within NVIDIA GPU and networking ecosystems, delivering high-throughput and low-latency data movement over intra-node NVLink/PCIe interconnects and inter-node InfiniBand (IB)/RoCE fabrics. RCCL extends the NCCL design to AMD GPU ecosystems.

\subsection{Communication Compression} \label{subsec:comm_compr}

Communication compression accelerates distributed data exchange by reducing the volume of transmitted data. It is commonly categorized into lossless and lossy approaches, whose effectiveness is inherently data-dependent: achievable compression ratio and throughput vary with data distribution, while lossy methods additionally introduce accuracy trade-offs~\cite{tao_2017}. Lossless compression (e.g., DietGPU~\cite{dietgpu}, LZ4~\cite{lz4}, and MANS~\cite{huang2025mans}) preserves numerical fidelity but typically provides limited compression ratios and throughput, making it suitable for precision-sensitive applications under network bandwidth constraints~\cite{dietgpu}. In contrast, lossy compression (e.g., SZ~\cite{liang_2022}, SDP4Bit~\cite{jia2024sdp4bit}, and PRISM~\cite{lu2026prism}) trades numerical precision for substantially higher compression ratios and throughput, and is therefore widely adopted in error-tolerant applications and high-speed network environments~\cite{jia2024sdp4bit}. Prior studies have shown that lossy communication compression can significantly improve distributed application performance with negligible accuracy degradation~\cite{alistarh_2017}. 

To facilitate practical deployment, communication compression has been integrated into collective communication libraries for scientific computing and AI workloads. In scientific computing, ZCCL~\cite{Huang2025ZCCLSI} incorporates the SZ lossy compressor into MPI and co-designs compression and communication pipelines, while gZCCL~\cite{gzccl2024} extends the design to GPUs. For AI workloads, UCCL-ZIP~\cite{ma2026ucclzip} integrates the DietGPU lossless compressor~\cite{dietgpu} into UCCL~\cite{mao2025uccl} via pipelining and operator fusion; COCCL~\cite{liucoccl2026} supports customizable compression operators on NCCL; NCCLZ~\cite{wang2026ncclz} combines quantization with lossless entropy coding; and ZipCCL~\cite{lin2026zipccl} optimizes compression for Mixture-of-Experts (MoE) training. These libraries primarily target either scientific computing or AI training, with limited support for AI-for-Science applications. We benchmark them systematically in Sec.~\ref{sec:evaluation}.

\subsection{Limitations of Existing Benchmarks} 
\label{subsec: limitations}

Several benchmarks partially address the evaluation of compression-aware collective communication, as summarized in Tab.~\ref{tab:comparison}. Vendor benchmarks, including NVIDIA NCCL Tests~\cite{nccltests}, AMD RCCL Perftest~\cite{rccltests}, and Intel MPI Benchmarks~\cite{imb}, measure backend-specific collective latency and bandwidth.CommBench~\cite{commbench2024} benchmarks realistic communication behavior across MPI, NCCL, RCCL, and OneCCL on hierarchical HPC systems. OSU Micro-Benchmarks (OMB)~\cite{graham2005osu} provide portable MPI communication benchmarks, while OMB-Compr~\cite{ombcompr2025} extends OMB with ZFP-based compressed collectives and reports overall latency and reconstruction error. Despite their contributions, existing benchmarks share several limitations in evaluating compression-aware communication:

\begin{table}[t]
\centering
\caption{Comparison of existing benchmarks with CC-Bench. Performance Profiling: coarse- and fine-grained metrics; Flexibility and Extensibility: communication backends, datasets, and quality metrics switching and integration; Workload Characteristics: real-world datasets and computation interference.}
\label{tab:comparison}
\scriptsize
\setlength{\tabcolsep}{4.5pt}
\resizebox{\linewidth}{!}{
\begin{tabular}{lcccc}
\toprule
\multirow{2}{*}{\textbf{Benchmark}} & \textbf{Performance} & \multirow{2}{*}{\textbf{Flexibility}} & \multirow{2}{*}{\textbf{Extensibility}} & \textbf{Workload} \\
& \textbf{Profiling} & & & \textbf{Characteristics} \\
\midrule
Vendor (N/R/I) & $\bigcirc$ & $\times$ & $\times$ & $\times$  \\
CommBench & $\bigcirc$ & $\checkmark$ & $\times$ & $\bigcirc$ \\
OMB & $\bigcirc$ & $\checkmark$ & $\times$ & $\times$ \\
OMB-Compr & $\bigcirc$ & $\bigcirc$ & $\times$ & $\bigcirc$ \\
\midrule
\rowcolor{blue!12}
\textbf{CC-Bench (Ours)} & $\checkmark$ & $\checkmark$ & $\checkmark$ & $\checkmark$  \\
\bottomrule
\end{tabular}
}
\\[4pt]
\flushleft\scriptsize $\checkmark$:~Achieved; \hspace{6pt} $\times$:~Not~supported; \hspace{6pt} $\bigcirc$:~Partial
\end{table}

\textbf{Limited Extensibility.} Existing benchmarks provide limited flexibility in switching communication libraries, datasets, and error metrics. Most either support only a single communication library or require recompilation to switch among supported libraries, while offering no mechanism for integrating new libraries. Moreover, since compression introduces data dependencies, benchmarking must account for diverse datasets and corresponding error metrics; however, prior work lacks support for customizable dataset and metric evaluation.

\textbf{Incomplete Performance Analysis.} Existing benchmarks provide only coarse-grained latency across message sizes, lacking fine-grained profiling of communication/compression overhead breakdown, domain-specific metrics and hardware utilization. This limits the ability to identify performance bottlenecks in communication compression, resulting in a gap between benchmarked and end-to-end application performance.

\textbf{Lack of Datasets.} Existing benchmarks largely overlook diverse real-world application datasets despite their importance and limited availability. This restricts the evaluation of communication compression across heterogeneous data characteristics. Consequently, users must manually curate application-specific datasets for realistic benchmarking, substantially increasing evaluation cost.

\section{Methodology of CC-Bench} \label{sec:methodology}
We present our benchmark goals (Sec.~\ref{subsec:bench_goals}) and design principles (Sec.~\ref{subsec:design_prin}).

\subsection{Benchmark Goals}  \label{subsec:bench_goals}
\label{goals}

CC-Bench bridges the gap left by prior work (Tab.~\ref{tab:comparison}) as a practical, diagnostic tool for compression performance in data-dependent, application-relevant settings. It answers three questions: (1) What latency–throughput–accuracy trade-offs arise across compressor–backend combinations, and how do they vary with workload characteristics? (2) Where do bottlenecks reside once compression is deployed—in codec throughput, network bandwidth, or interference-induced contention? (3) Do conclusions from isolated micro-benchmarks hold under realistic conditions with real data, computation interference, and application-specific accuracy requirements? We structure the evaluation along four dimensions, each tied to one of these goals.

\textbf{Cross‑stack Sensitivity}. We sweep a matrix of compressors and communication backends, measuring latency, throughput, and error metrics such as Mean Absolute Error (MAE). This dimension targets design‑space coverage: by exposing performance and accuracy variations across combinations, it helps users identify deviation‑performance trade-offs and optimal stack for a given workload class without manual benchmarking.

\textbf{Fine-grained Phase Decomposition}. We decompose each collective call into compress, send, recv, and decompress stages using lightweight runtime instrumentation. This dimension targets bottleneck localization: by revealing whether the dominant cost lies in compression throughput, network transfer, codec or insufficient overlap. It guides optimization efforts to the right component.

\textbf{Hardware‑level Introspection}. To address intractable conditions when function decomposition is unavailable, we enable collection of hardware utilization timelines (GPU SM/DRAM, NIC bandwidth, PCIe throughput) synchronized with operation call timeline. This dimension targets resource contention diagnosis: by exposing whether throughput is bounded by computation, memory bandwidth, or interconnect saturation, users can estimate performance bottlenecks when phase decomposition is unavailable.

\textbf{Realistic Workload Conditioning}. We inject real‑world datasets (e.g., scientific datasets and LLM communication traces) and computation interference profiles into the benchmark. This dimension targets operational realism: it quantifies how compression behaves under conditions resembling production deployments, including interference from co‑located workloads, unlike micro‑benchmark settings where such effects are invisible.

Together, these four dimensions provide a systematic evaluation methodology—from macro-level stack comparison down to micro-level resource attribution—that captures the multifaceted performance profile of communication compression in a single, configurable benchmark suite.

\subsection{Design Principles} \label{subsec:design_prin}

In order to achieve our goals for evaluation, CC-Bench is motivated to be lightweight, application-oriented and heterogeneously flexible. It is designed around five principles that directly address the limitations identified in Sec.~\ref{subsec: limitations}.

\textbf{Extensible and Wide-coverage Design.} CC-Bench addresses the limited backend and compressor coverage of existing benchmarks through a unified abstraction layer that decouples communication, compression, and profiling into independently swappable components. Users can integrate any compressor into any communication backend via uniform interfaces. The suite also provides a configurable dataset loader and a lightweight LD\_PRELOAD-based injection system that requires no recompilation, enabling users to easily sweep across backends, compressors, datasets, and runtime parameters with minimal effort.

\textbf{Full-stack Bottleneck Decomposition.} To move beyond coarse-grained latency and throughput, CC-Bench exposes where time is actually spent. On the software side, wrapper-bassed profiling isolates compress, decompress, send, recv, and overlap overhead. On the hardware side, daemons concurrently capture utilization metrics (CPU, GPU, memory, NIC), revealing whether the bottleneck lies in computation, memory, or communication. This dual-layer approach diagnoses contention and overlap inefficiencies that aggregate metrics alone cannot expose, directly addressing the limitations of isolated communication benchmarks.

\textbf{Data‑dependent Accuracy Assessment.} Recognizing compression is inherently data‑dependent, CC‑Bench provides a pluggable metric framework supporting domain‑specific criteria. To facilitate evaluation across diverse data modalities, the suite ships with representative datasets — including scientific fields and prevalent LLM traces — alongside domain‑appropriate metrics such as PSNR for climate fields and SSIM or error quantiles for LLM tensors. This enables systematic characterization of how data modality and accuracy requirements reshape compression‑quality trade-offs.

\section{Design of CC-Bench} \label{sec:design}
In this section, we first give an overview of our benchmark framework(Sec.~\ref{overview}). Then, we present the details of each component in our benchmark framework and supported plugins(Sec.~\ref{design_compnent}--\ref{coverage}). Finally, we provide some examples of how to use our benchmark(Sec.~\ref{sec:usage}).

\begin{figure}[t!]
\centering

\includegraphics[width=\linewidth]{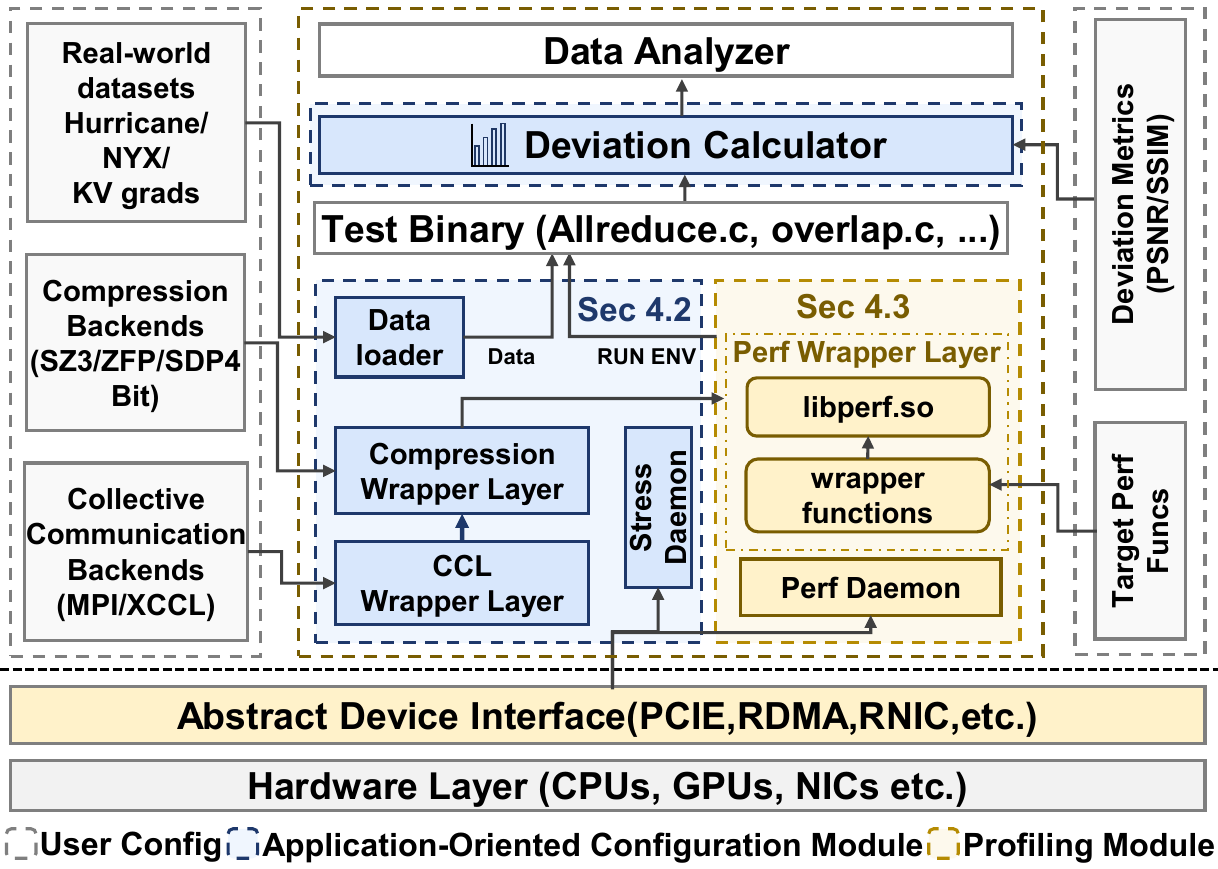}
\caption{CC-Bench architecture. The three swappable preload layers and daemons form the runtime for the test binaries.}

\label{fig:architecture}
\end{figure}

\subsection{Overview}\label{overview}

\textbf{Architecture.} 
Fig.~\ref{fig:architecture} illustrates CC-Bench's architecture, which consists of an application-oriented configuration module and a full-stack profiling module. The former uses communication and compression wrappers, a configurable dataset loader with pluggable deviation metrics, and a stress daemon to emulate realistic application environments. The latter adds a function-profiling wrapper and performance daemons whose hardware samples and traces are post-processed by the data analyzer. The benchmark extends to more nodes simply by editing the host list.

\begin{figure*}[h]
\centering
\includegraphics[width=1.0\linewidth]{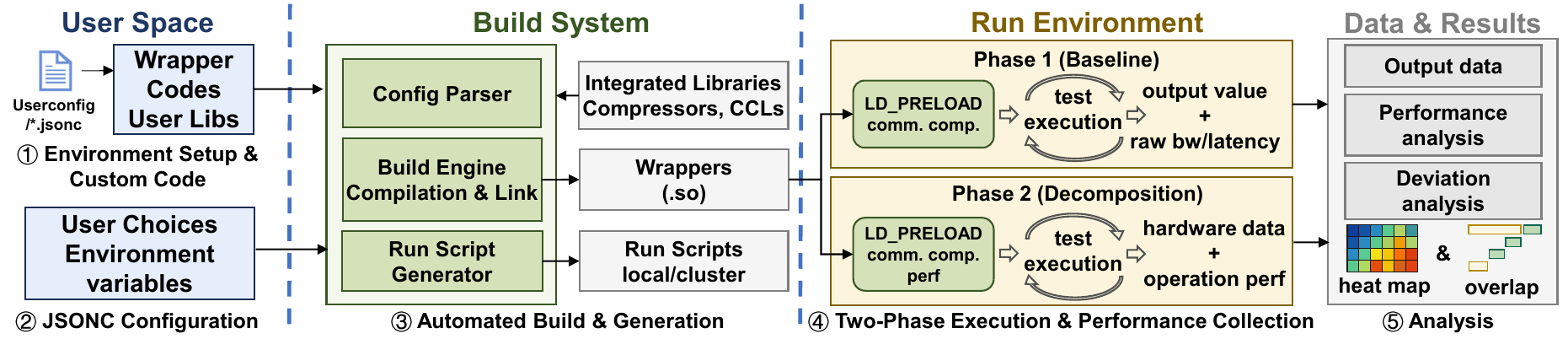}
\caption{CC-Bench workflow: \ding{172} Users configure benchmarks through JSONC files and \ding{173} provide custom wrapper code. \ding{174} The build stage compiles all wrappers
and generates execution scripts; \ding{175} the two-phase run collects performance and profiling data; \ding{176} the analysis stage produces multi-metric reports.}
\vspace{-4mm}
\label{fig:pipeline}
\end{figure*}

\textbf{Workflow.}
Fig.~\ref{fig:pipeline} exhibits CC-Bench's five-stage (Config, Customize, Build, Run, Analyze) workflow. \ding{172} The Configuration stage centralizes all parameters in modular JSONC files. \ding{173} Users provide wrapper code for libraries with inconsistent interfaces, forming the three wrapper layers. \ding{174} A single build script parses configurations, compiles wrappers, links backends, and emits a scheduler-aware run script. \ding{175} The run phase activates daemons and executes in two phases for precise latency and breakdown measurement. \ding{176} Analysis instruments compress/send/recv/decompress stages with hardware monitoring, from CSV metrics to overlap-efficiency and hardware-utilization heatmaps.

\subsection{Application-Oriented Configuration Module} \label{design_compnent}

\textbf{Communication Wrapper.} When target communication backends provide non-standard entry functions, this layer intercepts standard collective APIs (e.g., \texttt{MPI\_AllReduce}, \texttt{nccl\_AllReduce}) and reimplements them using compressed send/recv primitives or delegates to compression-aware libraries (e.g. ZCCL, UCCL-Zip) as shown in Fig.~\ref{fig:example_codes}(a), where we provide an instance of integrating ZCCL functions into standard MPI collectives. It maintains collective semantics while operating on compressed data, remaining compressor-agnostic.

\textbf{Compression Wrapper.} Similar to the previous layer, this wrapper allows weak-symbol defaults for compression entry points such as \texttt{compress}/\texttt{decompress} to be overridden by user-supplied codec shared libraries. This design enables users to intuitively substitute compression for certain backends, even when the backend has a built-in or hardcoded compressor. Our bench currently ships with wrappers for SZ3, ZFP for ZCCL backend and DietGPU, SDP4Bit wrappers for COCCL collectives.

\begin{figure}[t!]
  \centering
  \includegraphics[width=1.0\linewidth]{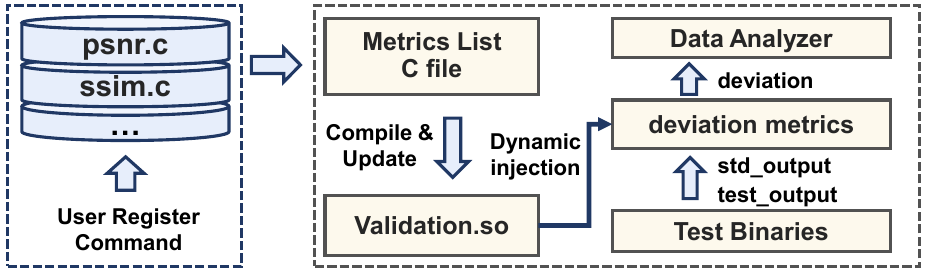}
  \caption{A user-friendly, lightweight design for dynamically plugging deviation metrics into the evaluation workflow.}
  \label{fig:design1}
  \end{figure}

\textbf{Dynamic Deviation Calculation.}
A dynamic deviation metric calculator ensures wide coverage across usage fields. As illustrated in Fig.~\ref{fig:design1}, users can implement custom metric functions (e.g., SSIM or domain-specific error measures) following a predefined template (see Fig.~\ref{fig:example_codes}(c)) and register them via a script run which automatically compiles them into a shared library. The framework then dynamically loads this library at runtime, seamlessly injecting the user-defined logic into the evaluation workflow without modifying core code. This design enables flexible, extensible metric integration across diverse application domains.

\begin{figure}[t!]
  \centering
  \includegraphics[width=1.0\linewidth]{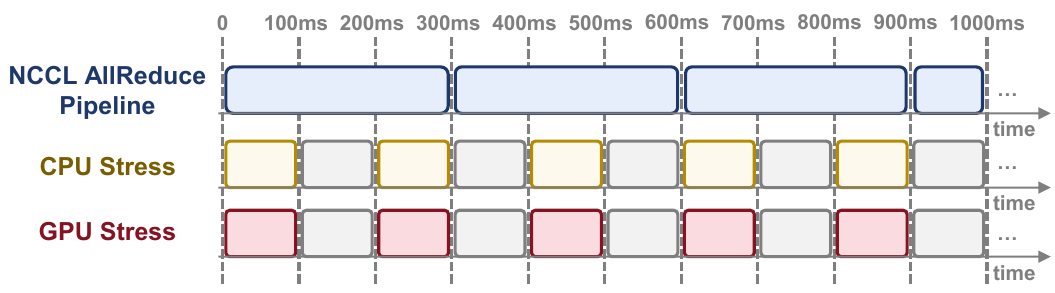}
  \caption{Overview of the stress daemon working process, intermittently spinning for a given portion of loops in the background.}
  \label{fig:design2}
  \end{figure}

\textbf{Stress Daemons.} In real-world HPC and AI clusters, background interference rarely manifests as sustained, full-saturation load. Instead, it typically arises from co-located jobs, system daemons, periodic synchronization routines, or network background traffic--all of which exhibit intermittent, bursty patterns with alternating active and idle phases. As illustrated in Fig.~\ref{fig:design2}, our stress daemons emulate such realistic contention through a configurable duty-cycle mechanism: they repeatedly execute lightweight compute loops during busy phases and remain idle during rest phases, with the busy/idle ratio specified by the user. We deliberately approximate interference with stress daemons rather than embedding full application stacks, since the latter would conflict with CC-Bench's lightweight, portable design.

\subsection{Full-stack Profiling Module}

\textbf{Profiling Wrapper.} The final layer of interception before function call. By bracketing the target function, users can trace per-call timestamps, message sizes, and compression ratios by utilizing the perf notedown helpers provided by our bench as shown in Fig.~\ref{fig:example_codes}(b). Activated only during decomposition phases, it provides traces which can further reveal macro-level information including overlap .

\textbf{Auxiliary Tests.} Apart from collective operations tests, The pingpong test records point-to-point latency/bandwidth across both intra-node and inter-node configurations. Given the linear or near-linear relationship between message size and latency, pingpong results enable interpolation-based estimation of non-blocking communication overhead in downstream tests. The overlap test measures computation-communication interference by dispatching compute and communication kernels concurrently on two GPU streams (or CPU threads) and comparing against serial execution. While the overlap test does not directly feed into other tests, it provides critical insight into how seemingly independent workloads interfere with each other when co-located.

\textbf{Perf Daemons.} Poll hardware counters (CPU, GPU, InfiniBand, PCIe etc.) and log timestamped samples. Similar to the deviation metrics system, they are plugged into the framework through a unified registration and dynamic loading mechanism, enabling easy extension. Sampling runs at a low frequency in a separate process and thus imposes negligible overhead on the measured communication path.

\textbf{Analysis Layer.} Consumes profiling wrapper traces and hardware samples to produce multi-dimensional performance views. Three statistical tools  serve different granularities: \texttt{draw\_hardware} plots hardware-counter heatmaps from daemon CSVs and supports time-range filtering, per-metric selection, and node isolation.\texttt{draw\_perf\_function} draws per-rank function-trace heatmaps from function traces of a specific rank with the chosen metrics (e.g. duration, compression ratio) and an aggregate  mode. \texttt{function\_analyzer} classifies calls into communication vs.\ compression, estimates async durations via pingpong interpolation, and reports overlap efficiency and idle time per phase. Together they
trace a bottleneck from macro utilization down to per-call overlap. Overall, these components expose overlap efficiency, per-function latency, hardware utilization, and their inter-dependencies.

\subsection{Benchmark Coverage} \label{coverage}

CC-Bench covers the main degrees of freedom needed to evaluate compression-enhanced collective communication. Our supported collective operations include Broadcast, Reduce, AllReduce, Scatter, Gather, AllGather, AlltoAll, ReduceScatter and vector collectives, enabling measurement of collective latency/bandwidth, point-to-point baselines, and slowdown under communication-computation overlap. Our currently integrated compression wrappers span CPU and GPU compressors such as SZ3~\cite{liang_2022}, ZFP~\cite{zfp2019error,lindstrom2014fixed}, DietGPU~\cite{dietgpu}, SDP4Bit~\cite{jia2024sdp4bit}, TAH-Quant~\cite{he2025tah}, and additional COCCL plugins, covering error-bounded, fixed-rate, lossless, and quantization-based compression paths. In terms of communication backends, we compare uncompressed MPI/NCCL baselines against compression-aware libraries including ZCCL~\cite{gzccl2024}, COCCL~\cite{liucoccl2026}, and UCCL-Zip~\cite{ma2026ucclzip} under a unified benchmark interface. Our data modality is highlighted: we capture data-dependent compression behavior across scientific datasets (Hurricane ISABEL~\cite{ieeevis2004isabel}, NYX~\cite{almgren2013nyx}, CESM-ATM~\cite{sdrbench21}), LLM KV-cache traces, Llama3~8B activation values, weights and gradients, spanning scientific computing, LLM inference/training, and controlled synthetic distributions. Finally, our variety of evaluation metrics link performance, fidelity, and profiling breakdowns to workload-specific constraints through latency, throughput, compression quality metrics, hardware traces, phase breakdowns, and configurable rank-offset schemes.

\subsection{Benchmark Usage} \label{sec:usage}

\begin{figure}[t!]
  \centering
  \includegraphics[width=1\linewidth]{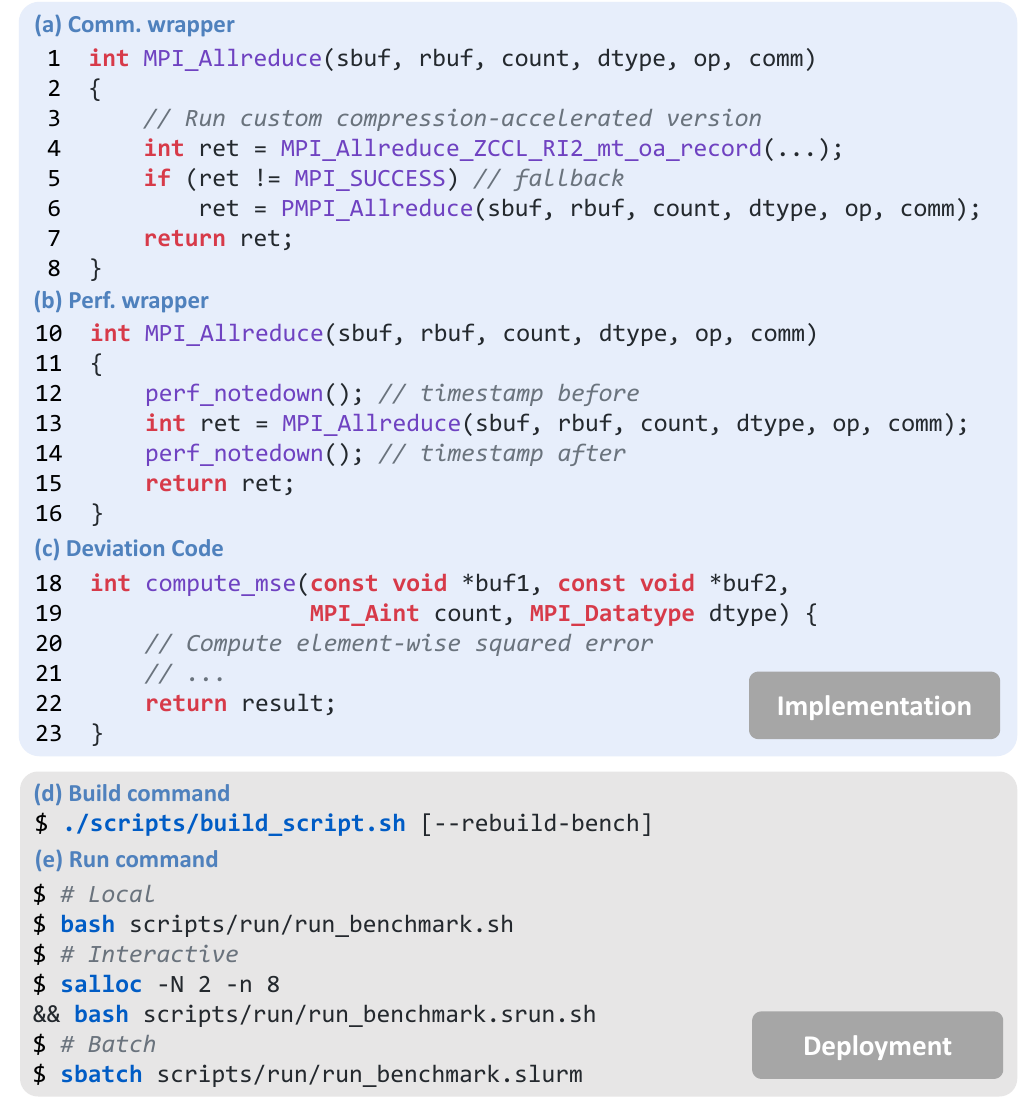}
  \caption{Example codes for deploying CC-Bench. Blue denotes custom C code templates, while gray denotes bash commands for setup and run.}
  \label{fig:example_codes}
\end{figure}

Fig.~\ref{fig:pipeline} shows the overall CC-Bench workflow: users configure benchmark inputs, customize optional extensions, build the run environment, execute tests, and analyze outputs. Fig.~\ref{fig:example_codes} complements this workflow by giving concrete examples of user-facing code formats and commands.

\textbf{Setup.} Setup corresponds to the config, customize, and build stages in Fig.~\ref{fig:pipeline}. Users flexibly select the execution environment, communication library, compressor, dataset, message sizes, background load, and accuracy metrics through modular JSONC configuration files. If a backend, profiler, daemon, or metric is not built in, users provide the lightweight extension formats shown in Fig.~\ref{fig:example_codes}(a)--(c). The build command in Fig.~\ref{fig:example_codes}(d) then compiles the selected extensions and emits a self-contained run script.

\textbf{Run \& Analysis.}
Run and analysis corresponds to the final two stages in Fig.~\ref{fig:pipeline}. Users execute the generated script as in Fig.~\ref{fig:example_codes}(e), producing latency, throughput, deviation metrics, per-rank traces, and hardware logs. Post-processing then reports phase breakdowns, overlap efficiency, and visualizations; hardware-utilization heatmaps provide a complementary bottleneck view when function-level interception is unavailable.


\section{Evaluation}
\label{sec:evaluation}


We evaluate CC-Bench as a diagnostic benchmark for compression-enhanced collective communication. Using AllReduce as the representative primitive, the evaluation process aims to answer the three questions crucial to resolving current limitations (see Sec.~\ref{goals}).

\begin{table}[b!] 
      \centering
      \caption{Quality metrics for CPU and GPU backend-compressor configurations. CPU rows report PSNR (dB, higher is better) and compression rate; GPU rows report cosine similarity(CosSim), relative error(RE), and compression rate(CR).}
      \label{tab:quality_metrics}
      \scriptsize
      
      \renewcommand{\arraystretch}{0.85}
      
      \begin{tabular*}{0.92\linewidth}{@{\extracolsep{\fill}}llccc@{}}
          \toprule
          \multicolumn{5}{c}{\textbf{(a) CPU backend-compressor}} \\
          \midrule
          \textbf{Backend} & \textbf{Metric} & \textbf{Hurricane} & \textbf{CESM} & \textbf{NYX} \\
          \midrule
          \multirow{2}{*}{ZCCL+SZx} & PSNR & 66.6 & 56.2 & -127.3 \\
          & CR & 0.104 & 0.104 & 1.026 \\
          \addlinespace
          \multirow{2}{*}{ZCCL+SZ3} & PSNR & 66.3 & 75.1 & -12.7 \\
          & CR & 0.117 & 1.033 & -- \\
          \bottomrule
      \end{tabular*}
      
      
      \begin{tabular*}{0.92\linewidth}{@{\extracolsep{\fill}}llcccc@{}}
          \toprule
          \multicolumn{6}{c}{\textbf{(b) GPU backend-compressor}} \\
          \midrule
          \textbf{Backend} & \textbf{Metric} & \textbf{Act.} & \textbf{Grad.} & \textbf{Weight} & \textbf{KV} \\
          \midrule
          \multirow{3}{*}{COCCL+SDP4Bit} & CosSim & 0.992 & 0.992 & 0.999 & 0.991 \\
          & RE & 86.710 & 0.669 & 0.903 & 3.85e4\\
          & CR & 0.133 & 0.133 & 0.133 & 0.133 \\
          \addlinespace
          \multirow{3}{*}{COCCL+TAH-Quant} & CosSim & 0.414 & 0.452 & 0.410 & 0.427 \\
          & RE & 6.81e5 & 19.1 & 8.39 & 6.96e6 \\
          & CR & 0.500 & 0.500 & 0.500 & 0.500 \\
          \addlinespace
          \multirow{3}{*}{UCCL+DietGPU} & CosSim & 1.000 & 1.000 & 1.000 & 1.000 \\
          & RE & 0 & 0 & 0 & 0 \\
          & CR & 0.846 & 0.851 & 0.844 & 0.855 \\
          \bottomrule
      \end{tabular*}
\end{table}
\subsection{Basic Benchmark Setup}

\textbf{Platform.} We evaluate CPU benchmarks on a cluster with two Intel Xeon Platinum 8358P CPUs (32 cores each) per node, using Open MPI 4.1.4 and 200G InfiniBand, with 16 tasks per node across 16 nodes. For GPU benchmarks, we use a GPU cluster with two AMD EPYC 7402 CPUs (24 cores each) and eight NVIDIA A800 SXM4 80~GB GPUs per node, with CUDA 12.8, NCCL 2.27, and 8 tasks per node across 2 nodes.

\textbf{Benchmark Library.} We adopt plain MPI and NCCL as uncompressed baselines. Other setups including ZCCL+SZx, ZCCL+SZ3, UCCL+DietGPU, COCCL+SDP4Bit and COCCL+TAH-Quant stand for diverse compression-accelerated communication schemes.

\textbf{Datasets.} To capture the data-dependent behavior of communication compression, CC-Bench uses scientific datasets, LLM workload tensors, and synthetic distributions. Scientific inputs include Hurricane ISABEL atmospheric fields, NYX cosmology fields, and CESM-ATM climate fields, covering 2D/3D arrays, multi-field simulations, and different entropy characteristics~\cite{sdrbench21}. LLM inputs include simulated KV-cache traces for inference and Llama~3~8B activations, gradients, and weights for training-oriented communication.

\textbf{Quality Metrics.} For an original vector $x=\{x_i\}_{i=1}^{n}$ and reconstructed vector $\hat{x}=\{\hat{x}_i\}_{i=1}^{n}$, CC-Bench reports pointwise and application-oriented quality metrics: 1) Mean Absolute Error (MAE) = $\frac{1}{n}\sum_{i=1}^{n}|x_i-\hat{x}_i|$, 2) Mean Squared Error (MSE) = $\frac{1}{n}\sum_{i=1}^{n}(x_i-\hat{x}_i)^2$, 3) Peak Signal-to-Noise Ratio (PSNR) = $10·\log_{10}\left(\frac{\mathrm{max\_val}^2}{\mathrm{MSE}}\right)$, 4) Cosine Similarity (CosSim) = $\frac{\sum_i x_i\hat{x}_i}
{\sqrt{\sum_i x_i^2}\sqrt{\sum_i \hat{x}_i^2}}$, 5) Relative Error (RE) = $\frac{\| x - \hat{x} \|_2}{\| x \|_2}$, 6) Compression Rate (CR) = $\frac{\text{size}_{\text{compressed}}}{\text{size}_{\text{original}}}$.







Performance and profiling results include latency(L), effective bandwidth(BW), slowdown(S) =  $L_{\mathrm{load}}/L_{\mathrm{baseline}}$, phase duration, SM and memory bandwidth utilization.

\subsection{Quality Evaluation}\label{exp2}

CC-Bench treats compression quality as workload-specific instead of a unified error metric. We present fidelity and compression ratios for typical CPU scientific datasets and GPU LLM workloads. CPU tests adopt PSNR to measure reconstruction quality across Hurricane ISABEL, CESM-ATM and NYX. GPU experiments leverage cosine similarity and relative error for Llama 3 8B activations, gradients, weights and KV-cache traces.

\begin{figure*}[htb]
\centering
\includegraphics[width=1.0\linewidth]{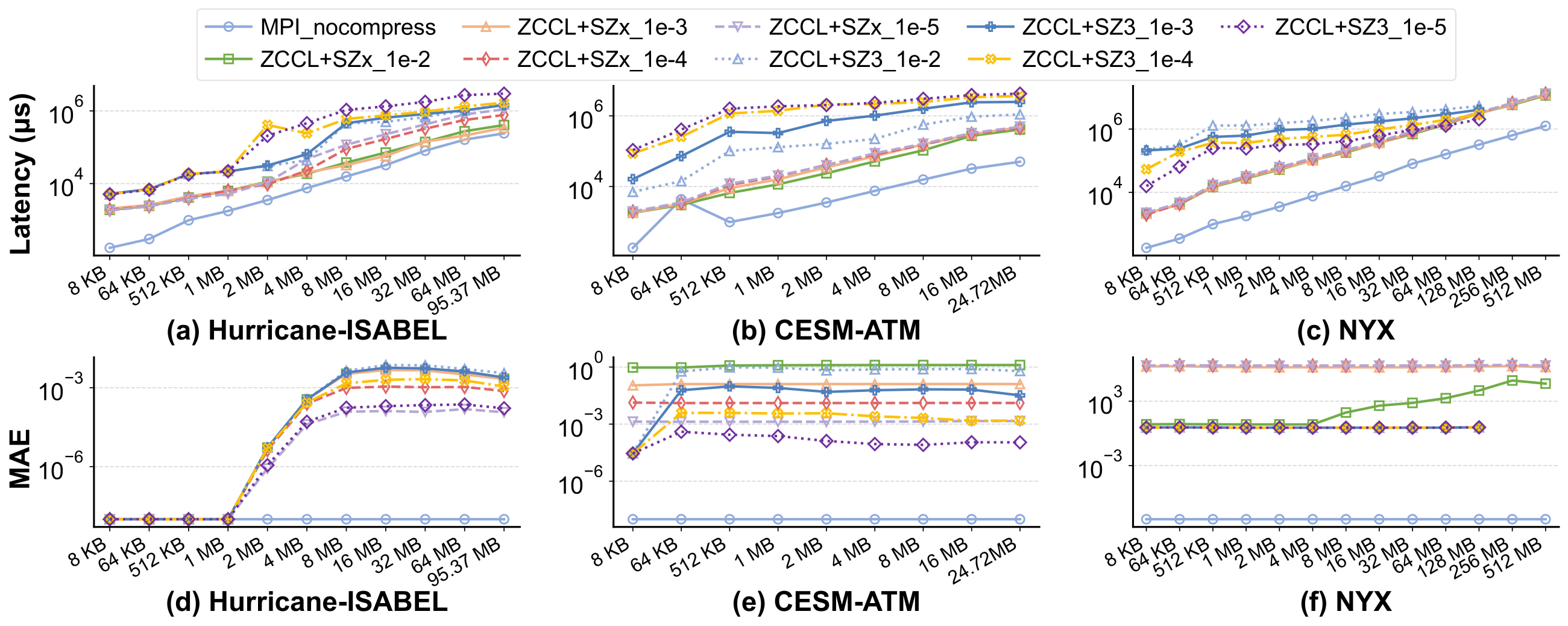}
\caption{CPU cross backend-compressor performance on scientific datasets. Latency incorporates overhead from all stages including codec and communication.}
\vspace{-2mm}
\label{fig:exp1_cpu}
\end{figure*}

\begin{figure*}[htb]
\centering
\includegraphics[width=1.0\linewidth]{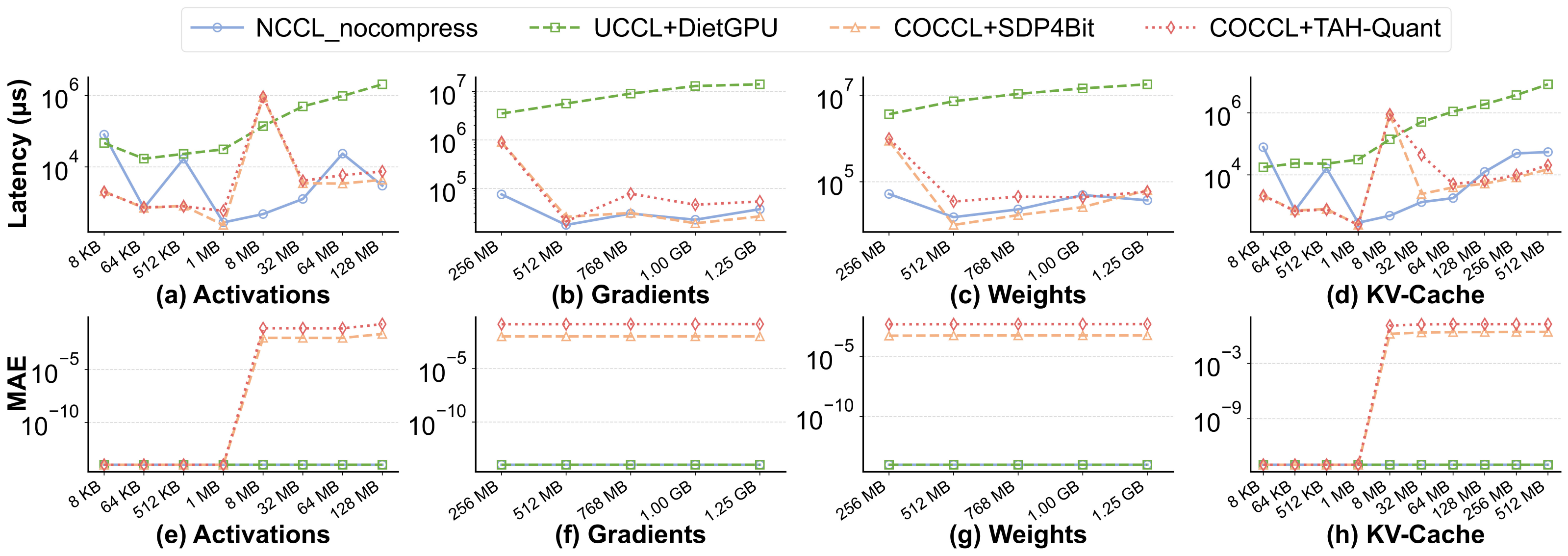}
\caption{GPU cross backend-compressor performance on LLM workloads. Latency incorporates overhead from all stages including codec, transfer and communication.}
\vspace{-4mm}
\label{fig:exp1_gpu}
\end{figure*}

Tab.~\ref{tab:quality_metrics}(a) shows that CPU compression quality is strongly dataset-dependent. For Hurricane ISABEL, ZCCL+SZx provides slightly higher PSNR than ZCCL+SZ3 (66.60 versus 66.31~dB) while also producing a lower compression rate (0.104 versus 0.117). For CESM-ATM, the quality-rate tradeoff reverses: ZCCL+SZ3 reaches 75.07~dB, but its compression rate is 1.033, whereas ZCCL+SZx reports 56.19~dB at a much lower rate of 0.104. NYX is a harder case for the tested compressors, with negative PSNR values and an SZx compression rate above one. 


\begin{observation}
    \textbf{Insight 1.1}: A lower compression rate does not necessarily mean that performance will be better. 
\end{observation}

The same holds on the GPU platfrom as observed in Tab.~\ref{tab:quality_metrics}(b).  The lossless UCCL-DietGPU achieves full accuracy with a lower compression ratio and uncompetitive thoughput,while interestingly SDP4bit significantly outperforms TAH-Quant in both cosine similarity and relative error while preserving a higher compression ratio of over 7.5. It can be also observed that COCCL+SDP4bit generally yield high relative errors, while cosine similarities remain promising, indicating that the reconstructed values are close to zero. These results clearly show that a compressor selected by a single dataset or metric can be potentially misleading.


\begin{observation}
    \textbf{Insight 1.2}: Failure on a single parameter does not necessarily indicate poor performance.
\end{observation}


\subsection{Performance Evaluation} 

 CC-Bench evaluates performance at three levels: overall latency for complete backend-compressor stacks(Sec.~\ref{exp1}), phase-level breakdown for bottleneck localization(Sec.~\ref{exp4}), and slowdown under background computation for realistic resource contention(Sec.~\ref{exp3}). This organization separates whether a compressed collective is faster, why it is faster or slower, and whether that conclusion remains valid when application kernels share the same hardware resources.

\subsubsection{Cross-combination Performance}\label{exp1}

This combinatorial experiment incorporates multiple backend-compressor stacks across two platforms, measuring each  as a deployable communication path  including  codec cost, transfer time, and reconstruction error. On the CPU cluster, we compare uncompressed MPI with ZCCL+SZx and ZCCL+SZ3 over NYX, Hurricane ISABEL, and CESM-ATM. On the GPU cluster, we compare uncompressed NCCL with UCCL+DietGPU, COCCL+SDP4Bit, and COCCL+TAH-Quant on KV-cache traces and Llama~3~8B activations, gradients, and weights. We record and observe the trade-offs between aggregate latencies and MAE of each combination.

Fig.~\ref{fig:exp1_cpu} reports CPU latency and MAE across the three scientific datasets. Compression is not a faster replacement for raw communication: MPI remains the lowest-latency choice in tested sizes, while compressed collectives add codec overhead that varies by dataset and error bound. For example, on Hurricane ISABEL, the best compressed latency at 64~MB with ZCCL+SZx (CR=$10^{-3}$) is 210,698.25~$\mu$s, compared with 161,494.36~$\mu$s for uncompressed MPI.

Fig.~\ref{fig:exp1_gpu} shows the corresponding GPU comparison. The LLM results are interpreted according to the natural message-size regimes of different tensor types,  following distinct message-size regimes: gradients and weights which represent data-parallel synchronization traffic are relatively large ($\geq$256MB), while KV-cache and activations which communicate in inference or tensor parallelism are small to medium ($\leq$512MB). We therefore sweep sizes accordingly.

For activation and KV-cache messages, COCCL quantization can sharply reduce latency at selected small sizes. At 8~KB, the best COCCL path achieves 36.36--40.33$\times$ speedup, and at 512~KB it achieves 21.34--21.45$\times$ speedup. The benefit is highly non-monotonic, however. At 8~MB, both COCCL paths exhibit a latency spike of 0.89--0.91~s while NCCL remains around 0.48~ms, indicating a size-specific overhead in the compressed path. For larger activation and KV-cache messages within the practical range, COCCL+SDP4Bit becomes competitive again, reaching 7.45$\times$ speedup on 512~MB activations and 5.98$\times$ and 3.60$\times$ speedups on 256~MB and 512~MB KV-cache messages, respectively.

\begin{figure}[t!]
\centering
\includegraphics[width=1.0\linewidth]{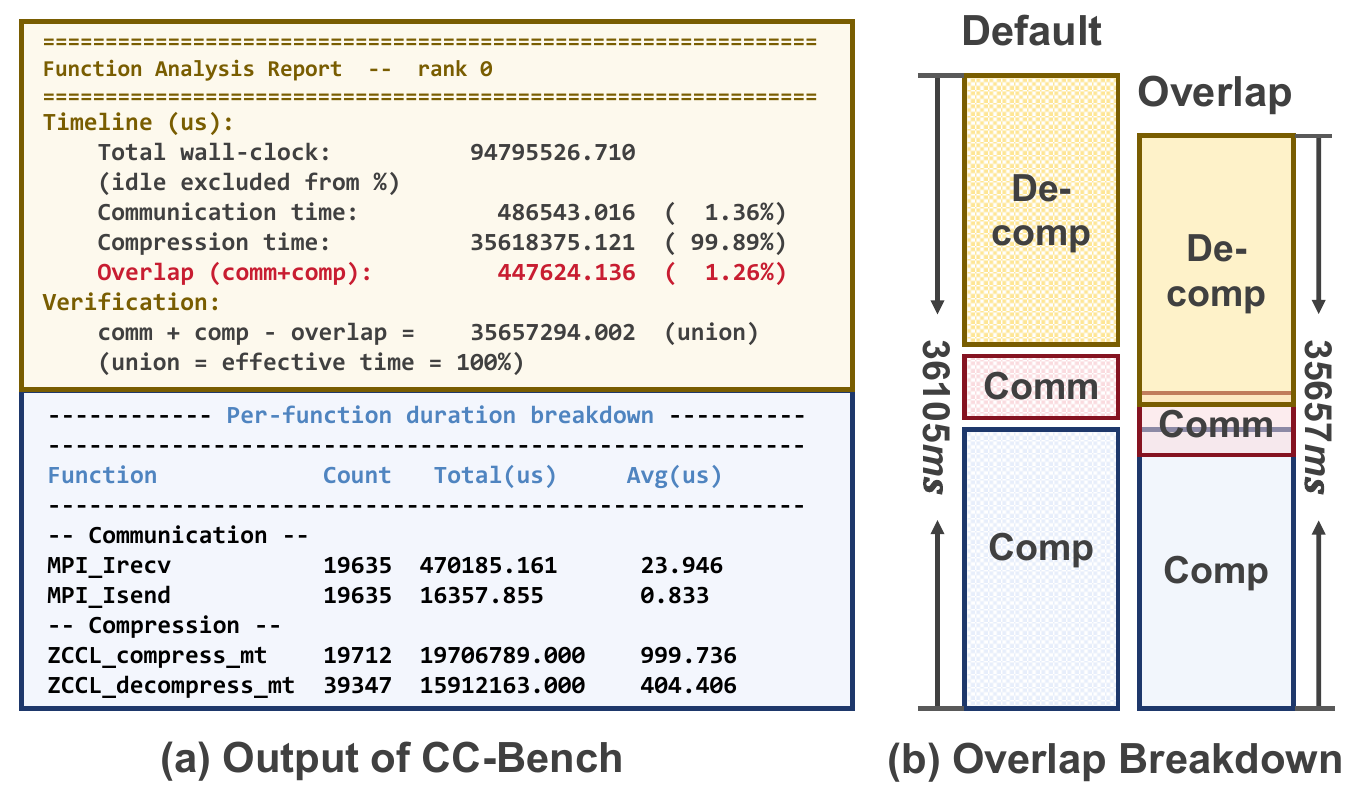}
\caption{The output of CC-Bench and CPU overlap breakdown for communication and compression phases.}
\label{fig:exp4_overlap}
\end{figure}

For gradients and weights, the large-message regime shows a more conservative but more deployment-relevant trend. On gradients, COCCL+SDP4Bit is slower than NCCL at the 256~MB boundary and remains close to parity at 768~MB, but becomes faster at 1.00~GB and 1.25~GB, with 1.18\(\times\) and 1.41\(\times\) speedups. On weights, COCCL+SDP4Bit outperforms NCCL at 512~MB, 768~MB, and 1.00~GB, with 1.52\(\times\), 1.36\(\times\), and 1.92\(\times\) speedups, respectively, although it is slower at 256~MB and 1.25~GB. Across all workloads, UCCL+DietGPU preserves zero MAE but is usually slower than NCCL, especially for medium and large messages, suggesting that lossless compression overhead and backend compatibility dominate the saved transfer time on this platform. Among the two lossy COCCL paths, SDP4Bit consistently has lower MAE than TAH-Quant, with MAE around \(1.04\times10^{-2}\) on gradients, \(5.06\times10^{-4}\)--\(5.37\times10^{-4}\) on weights, up to \(3.34\times10^{-2}\) on activations, and 1.63--2.64 on KV-cache messages of at least 8~MB. TAH-Quant is occasionally faster at some tiny-message points, but it consistently incurs larger reconstruction error.


\begin{observation}
    \textbf{Insight 2.1}: The selection of a compressor or backend should only be made once the quality gain, data reduction, execution overhead and compatibility with hardware have been considered together.
\end{observation}

\subsubsection{Performance Breakdown} \label{exp4}


The performance-breakdown experiment explains where total time is spent. Here we demonstrate the per-operation breakdown for ZCCL+SZx, since CC-Bench  intercepts its communication functions such as MPI\_Isend and codec functions such as SZx calls with minimal effort using its customized profiling wrappers to. The post-run analyzer then separates compression, decompression, send/recv, wrapper overhead, and overlap time, as reported in Fig.~\ref{fig:exp4_overlap}. 

On the CPU platform, the overlap breakdown shows that communication is mostly hidden by the codec workflow rather than appearing as exposed transfer time. The profiled communication time is 486,543~$\mu$s, but only 38,919~$\mu$s remains after overlap analysis. In particular, 447,289~$\mu$s overlaps with compression and 335~$\mu$s overlaps with decompression, meaning that approximately 92.0\% of communication time is concurrent with codec work. The exposed timeline is therefore dominated by compression-only and decompression-only regions, which explains why optimizing only the communication backend or overlap mechanism would have limited impact for this configuration.

At a finer granularity, the analysis output also reveals notable asymmetries existing between operations of the same category. While compression incurs higher per-call latency, its lower invocation count (19,712 vs. 39,347 for decompress) results in comparable total execution time between the two phases. This suggests that the communication scheduler may benefit from prioritizing overlap of the more frequent decompression operations with `Irecv`, rather than treating compression and decompression symmetrically.

\begin{observation}
    \textbf{Insight 2.2}: Codecs dominate the ZCCL timeline, with communication largely hidden behind its operations -- decompression by volume, compression by weight.
\end{observation}


It is also worth noting that for many GPU backends such as UCCL-zip,  whose compression and communication entry points are less directly separable , CC-Bench can resort to reading hardware counters including SM utilization and network-device utilization over successive time windows. These time-resolved hardware states also provide complementary signals for locating bottlenecks through exposing overlap and underutilization of components. 
 \begin{figure}[t!]
 \centering
 \includegraphics[width=1.0\linewidth]{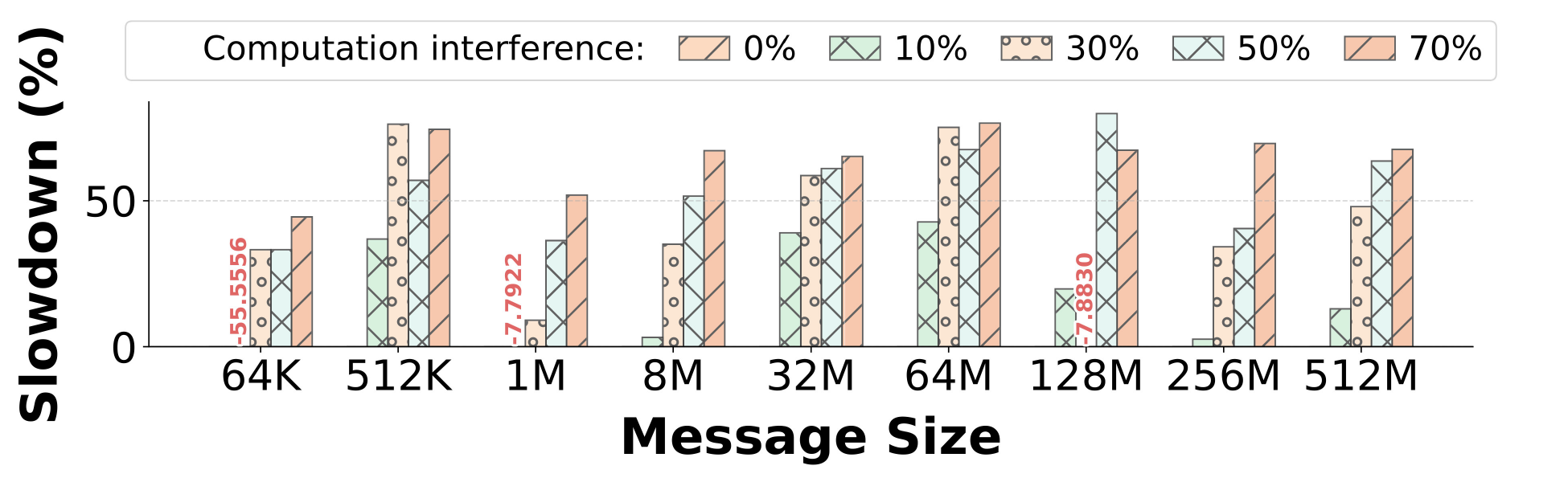}
 \caption{Slowdown of COCCL+SDP4Bit on KV-cache. Slowdown (\%) denotes latency increase relative to the no-load baseline; percentages indicate injected background load intensity.}
 \label{fig:exp3_Slowdown}
 \end{figure}

\subsubsection{Performance Slowdown}\label{exp3}

The slowdown experiment evaluates whether clean-environment performance holds when communication compression shares GPU resources with application computation. We use CC-Bench to inject controlled background load, evaluate COCCL+SDP4Bit on KV-cache traces, sweep GPU load from 0\% to 70\%, and record the resulting latency degradation.


As shown in Fig.~\ref{fig:exp3_Slowdown}, slowdown generally increases as GPU background load becomes heavier, but the effect is also message-size dependent. The largest-message and high-load cases consistently show substantial latency degradation, while very small messages and several intermediate-size points are non-monotonic. In particular, low background load can occasionally reduce measured latency, reflecting launch overhead, scheduling noise, and the sensitivity of short collectives to runtime variance. These results show that compressed-communication latency measured in isolation can overestimate practical speedup when compression kernels contend with application kernels.

\begin{observation}
    \textbf{Insight 2.3}: Background computation can erode compression-enhanced communication throughput.

\end{observation}

\subsection{Insight Discussion}

\textbf{Communication compression should be evaluated as a workload-conditioned communication path, not as an isolated codec property.}
The evaluation directly addresses the first question in Sec.~\ref{subsec:bench_goals} on how performance-accuracy trade-offs vary across stacks and data. As observed in Insight~1 and Insight~2.1, higher compression ratios improve communication efficiency at the cost of increased encoding overhead and potential fidelity loss. However, tradeoff trends shift substantially with workload characteristics---low-entropy HPC data tend to be compression-ratio-driven, while high-entropy training workloads are often accuracy-driven. Compression is therefore a workload-conditioned optimization, not a universal replacement; users should apply CC-Bench to full backend-compressor-data combinations rather than any single dimension in isolation.

\textbf{Performance bottleneck location requires explaining why a stack is fast or slow.}
The finer the decomposition of the overall operation, the more precisely users can locate what throttles macro efficiency. Insight~2.2 reveals through phase decomposition that the bottleneck for ZCCL is codec work, not communication itself. Following this logic, CC-Bench provides profiling coverage across the entire granularity spectrum---from coarse module-level (communication vs. compression) down to function-level (compress, decompress, send, recv), and further to function-internal parameters such as compression ratio or per-call latency. This coarse-to-fine breakdown lets users narrow the bottleneck from communication-versus-computation down to the specific operation and its internal characteristics. CC-Bench thus bridges clean micro-benchmarks and full-application profiling.

\textbf{CC-Bench also reflects end-to-end behavior by preserving context around communication.}
Full applications do not execute communication compression in isolation: LLM training and inference overlap collectives, compression kernels, and application kernels on the same GPU resources.  Insight~2.3 demonstrates that this interference is non-negligible by showing that shared GPU computation interference can substantially reduce compression-enhanced communication performance. CC-Bench approximates this setting by preserving the full communication path and injecting configurable GPU background load---for instance, 30\% to simulate a contended inference server, or 70\% to model training with colocated computation.


\section{Conclusion and Future Work} \label{sec:conclusion}

This paper presents CC-Bench, a lightweight benchmark suite for compression-enhanced collective communication. Its core contribution is twofold: a flexible and easy-to-use configuration workflow for execution environments, communication libraries, compressors, accuracy metrics, and real application datasets; and a profiling path that breaks benchmark runs into software phases and hardware-utilization evidence. Our evaluation shows that latency, throughput, and accuracy must be considered jointly, that bottlenecks shift with overlap and interference, and that isolated microbenchmark conclusions do not always hold under realistic conditions.

Future work extends CC-Bench along three directions. First, error propagation profiling tracks compression error variation across iterative compression-decompression steps in collectives. Second, the plugin framework integrates new compressors, hardware backends and I/O/power monitoring daemons. Third, lightweight proxies replay collective traces in mini-apps and training loops, bridging benchmark tests and practical application behaviors.

\section*{Acknowledgments}
This work was supported by the National Key Research and Development Program of China (Grant No. 2025YFB3003702), the National Natural Science Foundation of China (Grant No. T2125013), and the Innovation Funding of ICT, CAS (Grant No. E461050). The experiments were performed on the robotic AIScientist platform of Chinese Academy of Sciences.

\newpage
\bibliographystyle{IEEEtranS}
\bibliography{reference}

\newpage
\appendices
\section{Artifact Appendix}

\subsection{Abstract}

This artifact contains the full source code of CC-Bench, which is available at CCBench\_IISWC\_AE. This appendix gives instructions to regenerate the experiments described in section 5 of the paper.

\subsection{Artifact check-list (meta-information)}

{\small
\begin{itemize}
  \item {\bf Algorithm: } Lossy compression in collective communication
  \item {\bf Program: } C, CUDA
  \item {\bf Compilation: } gcc, mpicc, nvcc
  \item {\bf Data set: } Hurricane ISABEL, Llama~3~8B gradients.
  \item {\bf Run-time environment: } Linux; Open MPI; NCCL; CUDA
  \item {\bf Hardware: } CPU: 2$\times$Xeon Platinum 8358P per node $\times$ 2 nodes, 10G InfiniBand; GPU: 2$\times$EPYC 7402, 4$\times$A800 SXM4 80GB per node $\times$ 2 nodes, 200GbE+ Infiband;
  \item {\bf Metrics: } Compression Rate, MAE, MSE, PSNR, CosSim, Relative Error; latency, bandwidth, Slowdown Percentage
  \item {\bf Output: } terminal metric results, per-function traces
  \item {\bf Experiments: } domain-specific error(E1), quality--performance trade-off(E2), Interference slowdown(E3), overhead breakdown(E4)
  \item {\bf How much disk space required: } $\approx$25GB
  \item {\bf Time needed to prepare workflow: } $\approx$ 15 minutes
  \item {\bf Time needed to complete experiments: } $\approx$ 15 minutes
  \item {\bf Data licenses:  } MIT-License (Hurricane ISABEL)
  \item {\bf Workflow automation: } Semi (Shell script + manual run)
  \item {\bf Archived:  } \url{https://doi.org/10.5281/zenodo.21849825}  
\end{itemize}
}

\subsection{Description}

Our experiments coincide with the main contributions stated in our paper as follows:

\paragraph{C1} We introduce an \textbf{application-oriented configuration module} that decouples profiling logic from communication libraries, datasets, and error metrics.

\paragraph{C2} We develop a \textbf{full-stack profiling module} that combines function-level interception with hardware counter monitoring.

\paragraph{C3} We incorporate \textbf{built-in representative datasets} spanning scientific computing and LLM training/inference.

\begin{table}[h]
	\centering
	\label{tab:contrib-exp-map}
	\begin{tabular}{p{0.30\linewidth} p{0.20\linewidth} p{0.3\linewidth}}
		\toprule
		\textbf{Contribution} & \textbf{Supporting Artifact} & \textbf{Supporting Experiments} \\
		\midrule
		C1 & A & E1, E2  \\  
		C2 & A & E3, E4  \\  
		C3 & A & E1, E2  \\  
		\bottomrule
	\end{tabular}
\end{table}

\subsubsection{How to access}

The artifact is available at \url{https://github.com/konnyakucstdio/CCBench_IISWC_AE.git}. A frozen version is archived at Zenodo with DOI: \url{https://doi.org/10.5281/zenodo.21849825}.

\subsubsection{Hardware dependencies}

see Hardware section in A.2.

\subsubsection{Software dependencies}

\begin{itemize}
    \item \textbf{OS:} Ubuntu 22.04+(GPU) or CentOS7+(CPU)
    \item \textbf{Compiler:} GCC $\ge$ 9.0(GPU), 7.3.1(CPU); 
    \item \textbf{Python:} $\ge$ 3.8(GPU), 2.7(CPU)
    \item \textbf{MPI:} Open MPI 4.1.4
    \item \textbf{CUDA:} 12.8 with NCCL 2.27
    \item \textbf{Compression libraries:}SZx\textsuperscript{\href{https://github.com/szcompressor/SZx}{[1]}}, SDP4bit\textsuperscript{\href{https://github.com/ByteDance-Seed/SDP4Bit}{[2]}}
    \item \textbf{Communication libraries:} CoCCL\textsuperscript{\href{https://github.com/hpdps-group/coccl}{[3]}}, ZCCL\textsuperscript{\href{https://github.com/ZCCLorg/zccl}{[4]}}
\end{itemize}

\vspace{-4pt}
{

\noindent we recommend using our \textbf{locally revised libraries}\textsuperscript{\href{https://huggingface.co/datasets/konnyakucstdio/CCBench_IISWC_AE/tree/main/assets}{[5]}} due to bugs in the original version.
}

\subsubsection{Data sets}

\begin{itemize}
    \item \textbf{Hurricane:}  from \url{https://sdrbench.github.io/}.

    \item \textbf{Llama~3~8B:} gradients captured from runs of Llama~3~8B. This will be provided in our link\textsuperscript{\href{https://huggingface.co/datasets/konnyakucstdio/CCBench_IISWC_AE/tree/main/assets}{[5]}}.

\end{itemize}

\subsection{Installation}

Compulsory if running first time on new platform.

\noindent \circlednum{1} Clone the repository, datasets and libs:
\begin{lstlisting}[language=bash,linewidth=0.98\linewidth]
git clone https://github.com/konnyakucstdio/CCBench_IISWC_AE.git
cd CCBench_IISWC_AE
./scripts/download_data.sh
\end{lstlisting}

\noindent \circlednum{2} Set up the toolchain via \texttt{module load}. If modules are unavailable, manually export the environment variables following the example in \texttt{scripts/export\_env.sh}.
\begin{lstlisting}[language=bash,linewidth=0.98\linewidth]
# On GPU:
module load cuda/12.8 openmpi/4.1.5_cuda12.8 ucx/1.12.1_cuda12.8 nccl/2.27_cuda12.8
# On CPU:
module load compiler/dtk/22.04.2 compiler/devtoolset/7.3.1 mpi/hpcx/gcc-7.3.1
\end{lstlisting}

\noindent \circlednum{3} Set the \texttt{communication\_arch} field to \texttt{mpi} or \texttt{nccl} in \texttt{userconfig/config\_in\_jsonc/bench\_basic\_config}, then build:
\begin{lstlisting}[language=bash,linewidth=0.98\linewidth]
./scripts/build_script.sh --rebuild-bench
\end{lstlisting}

\subsection{Experiment workflow}

\paragraph{GPU platform}

To reproduce experiments E1 and E2:

\noindent \circlednum{0} Replace the \texttt{cross\_alloc\_nodes} in \texttt{userconfig/ config\_in\_jsonc/job\_config.jsonc} with the nodes available for run.

\noindent \circlednum{1} Build the benchmark
\begin{lstlisting}[language=bash,linewidth=0.98\linewidth]
./scripts/build_ae.sh 1 nccl
\end{lstlisting}

\noindent \circlednum{2} Run the benchmark:
\begin{lstlisting}[language=bash,linewidth=0.98\linewidth]
./scripts/run/run_benchmark.mpirun.sh
\end{lstlisting}

\noindent Likewise,To reproduce experiment E3:

\begin{lstlisting}[language=bash,linewidth=0.98\linewidth]
./scripts/build_ae.sh 3 nccl
./scripts/AE_E3.sh
\end{lstlisting}

\paragraph{CPU platform}

\noindent To reproduce E1, E2 and E4 on CPU:

\noindent \circlednum{0}\noindent Same as \circlednum{0} on GPU platform.

\noindent \circlednum{1} Build and prepare the pingpong baseline (all-in-one).:
\begin{lstlisting}[language=bash,linewidth=0.98\linewidth]
./scripts/build_ae.sh 4 mpi
\end{lstlisting}

\noindent \circlednum{2} Run the E1 \& E2 benchmark:
\begin{lstlisting}[language=bash,linewidth=0.98\linewidth]
./scripts/run/run_benchmark.mpirun.sh
\end{lstlisting}

\noindent \circlednum{3} For E4, analyze the function traces:
\begin{lstlisting}[language=bash,linewidth=0.98\linewidth]
python scripts/statistic/function_analyzer.py
\end{lstlisting}

\subsection{Evaluation and expected results}

Outputs for experiments can be found in dir \texttt{ccbench\_reports}. 
Experiments 1 and 2 should be like:
\begin{lstlisting}[language=bash,linewidth=0.98\linewidth]
Size: 536870912 bytes (134217728 elements)
User: avg=11100.49 us,min=11071.08 us,max=11135.39 us
Bandwidth: avg=   96.73 GB/s, min=   96.43 GB/s, max=   96.99 GB/s
Correct: NO
cos_sim: 9.813335e-01
mae: 3.090251e-04
...
\end{lstlisting}
{ This should be consistent with the results in Table 2, Figure 7 and Figure 8 in our paper. }

For experiment 3:

\begin{lstlisting}[language=bash,linewidth=0.98\linewidth]
=== AE_E3 summary: latency (us) / bandwidth (GB/s) per GPU stress level ===
stress%    avg_us      avg_GBps    vs-baseline
0%        3739.50    143.59      1.00x
10%        15485.43   34.69       0.24x
...
\end{lstlisting}
{ This should be similar to Figure 10 in our paper. }

For experiment 4:
\begin{lstlisting}[language=bash,linewidth=0.98\linewidth]
Timeline (us):
Total wall-clock:          20812851.140
Communication time:         7055756.000  ( 33.90%)
Compression time:           4272003.499  ( 20.53%)
Overlap (comm+comp):        4272003.499  ( 20.53%)
...
\end{lstlisting}
{ This should be similar to Figure 9 in our paper. }

\subsection{Experiment customization}

CC-Bench supports extensive customization through eight independent JSONC configuration files and implementing backend, compressor, profiling kernels and deviation metrics. Users can explore more by toggling all files under the userconfig directory.

\end{document}